\documentclass[twocolumn,prl,aps,superscriptaddress,showpacs]{revtex4}

\usepackage{dcolumn}
\usepackage{amsmath}
\usepackage{epsfig}

\newlength{\figwidth}
\preprint{DRAFT}

\begin{document}


\title{Resonant inelastic x-ray scattering study of charge excitations in
La$_2$CuO$_4$}
\author{Y. J. Kim}
\author{J. P. Hill}
\affiliation{Department of Physics, Brookhaven National Laboratory,
Upton, New York 11973}
\author{C. A. Burns}
\affiliation{Department of Physics, Western Michigan University,
Kalamazoo, Michigan 49008}
\author{S. Wakimoto}
\thanks{Current address: Department of Physics, University of Toronto,
Toronto, Ontario M5S 1A7, Canada}
\affiliation{Center for Materials Science and Engineering,
Massacusetts Institute of Technology, Cambridge, Massachusetts
02139}
\author{R. J. Birgeneau}
\affiliation{Department of Physics, University of Toronto,
Toronto, Ontario M5S 1A7, Canada}
\author{D. Casa}
\author{T. Gog}
\author{C. T. Venkataraman}
\affiliation{CMC-CAT, Advanced Photon Source, Argonne National
Laboratory, Argonne, Illinois 60439}

\date{\today}

\begin{abstract}

We report a resonant inelastic x-ray scattering study of the dispersion
relations of charge transfer excitations in insulating La$_2$CuO$_4$. These
data reveal two peaks, both of which show two-dimensional characteristics.
The lowest energy excitation has a gap energy of $\sim 2.2$ eV at the zone
center, and a dispersion of $\sim 1$ eV. The spectral weight of this mode
becomes dramatically smaller around ($\pi$, $\pi$). The second peak shows a
smaller dispersion ($\sim 0.5$ eV) with a zone-center energy of $\sim 3.9$
eV. We argue that these are both highly dispersive exciton modes damped by
the presence of the electron-hole continuum.

\end{abstract}

\pacs{74.25.Jb, 74.72.Dn, 78.70.Ck, 71.35.-y}

\maketitle


Understanding strongly correlated electron systems, such as the cuprate
superconductors, remains at the heart of much of current condensed matter
research. As a first step toward elucidating electron correlation effects
in these systems, it is important to study the behavior of elementary
excitations using various spectroscopic tools. For example, angle resolved
photoemission spectroscopy (ARPES) has become an indispensable probe for
studying the excitation spectrum of a single quasiparticle
\cite{Wells95,Pothuizen97}, while inelastic neutron scattering has been
invaluable in the investigation of low-energy collective modes ($1 \sim 100$
meV), such as phonons and magnons \cite{Birgeneau89}. However, to date only
limited information has been available on collective excitations at an
energy scale on the order of $\sim 1$ eV. This is unfortunate, since in this
energy range, the electron dynamics are governed directly by the various
hopping integrals and by the Coulomb interaction. Thus, an investigation of
the dispersion relations of such collective charge excitations would yield
invaluable information for any microscopic theory of charge dynamics in
the copper oxides.

In this Letter, we present a detailed study of the momentum-dependence of
the charge excitations in La$_2$CuO$_4$, utilizing resonant inelastic x-ray
scattering (RIXS). In the simplest picture of this insulating cuprate, the
ground state consists of one hole per copper ion (Cu$^{2+}$), and the
low-lying electronic excitations include excitons formed via a charge-transfer
(CT) process, in which charge is moved from the oxygen onto the copper
\cite{Simon96,Zhang98,Hanamura00}. Specifically, an electron-hole pair,
created by exciting an electron from the valence band -- the Zhang-Rice (ZR)
band \cite{Zhang88} -- to the conduction band across the CT gap, can form a
bound exciton state as a result of the Coulomb interaction. This CT exciton
is expected to have a large dispersion, since it has zero spin and can move
without disturbing the antiferromagnetic order of the copper oxide plane.
Consistent with this, our high-resolution measurements have enabled us to
identify an exciton-like feature at 2.2 eV with a dispersion of 1 eV.
In addition, a second peak at slightly higher energy is also observed. This
has a zone-center energy of 3.9 eV and a dispersion of 0.5 eV. Finally, we
also discuss a dramatic reduction in spectral weight of the CT exciton
observed near the ($\pi$ $\pi$) position.

The RIXS technique in the hard x-ray regime is a powerful new experimental method to
probe momentum-dependent collective excitations in condensed matter systems. As first
observed in NiO \cite{Kao96}, large enhancement of the inelastic x-ray scattering
signal is obtained when the incident x-ray energy is tuned near the transition metal
K-edge. To date, a number of insulating cuprates have been studied
\cite{Hill98,Abbamonte99,Hasan00,Hasan02}. While these previous studies provided
detailed information on the local resonance processes, they provided only qualitative
information on the momentum({\bf Q})-dependence of the CT excitations, mainly due to
poor signal-to-noise ratio. In the present work, we rectify this, studying in detail
the {\bf Q}-dependence of the RIXS spectrum in La$_2$CuO$_4$.


This work was carried out at the Advanced Photon Source on the undulator
beamline 9IDB. A double-bounce Si(111) monochromator and a Si(333)
channel-cut secondary monochromator was utilized.  A spherical, diced, Ge(733)
analyzer was used to obtain an overall energy resolution of 0.4 eV (FWHM).
The scattering plane was vertical and the polarization of the incident x-ray
was kept fixed along the {\bf c}-direction for all data reported here. The
single crystal sample of La$_2$CuO$_4$ was grown using the traveling solvent
floating zone method, and annealed to remove excess oxygen.  The crystal was
cut along the (100) plane and mounted on an aluminium sample holder at room
temperature. Throughout this paper, we use the tetragonal notation ($a
\approx b \approx 3.85 \AA $ along the Cu-O-Cu bond direction).


\begin{figure}
\begin{center}
\epsfig{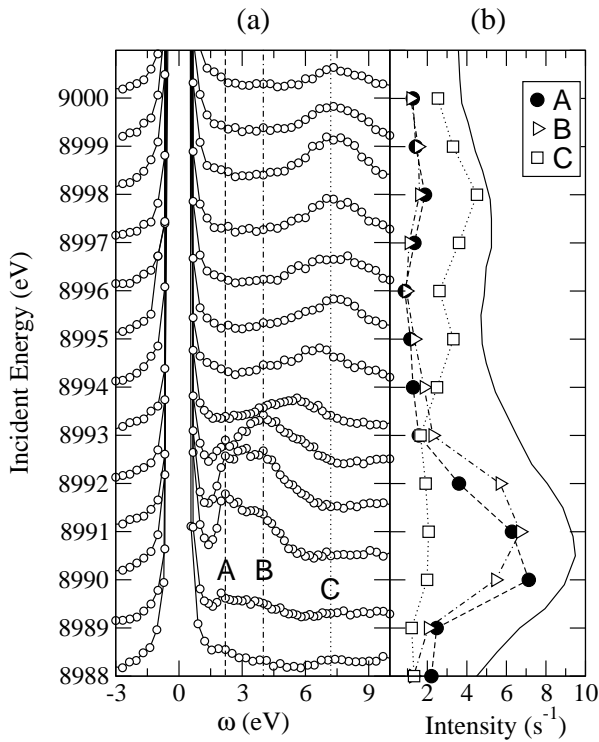}
\end{center}
\caption{
(a) Scattered intensity at the zone center, {\bf Q}=(3 0 0), plotted as a
function of energy transfer, $\omega$. The incident energy for each scan
can be read off from the vertical axis.  (b) Scattered intensity at the
three energy transfers, labeled A, B, and C in (a), as a function of
$E_i$. Also plotted is the x-ray absorption (in arbitrary units) as measured by 
monitoring the fluorescence yield (solid line).}
\label{fig1}
\end{figure}

In Fig.\ \ref{fig1}(a), we plot the incident energy ($E_i$) depedence of the
scattered intensity as a function of energy transfer ($\omega$) at a fixed
momentum transfer of {\bf Q}=(3 0 0), which corresponds to the Brillouin
zone (BZ) center. A resonant feature is observed between 2 eV and 4
eV for $E_i=8991$ eV, becoming weaker as the incident energy is
tuned away from resonance. This feature consists of two peaks, labeled A and
B, which show slightly different resonance behavior. In addition, there is
another feature (C) at 7.2 eV, which resonates around $E_i \approx 8998 $
eV. To show the resonance profile of these peaks, we plot the
scattered intensity as a function of $E_i$, with the energy transfer,
$\omega$, fixed at each excitation: $\omega_A=2.2$ eV, $\omega_B = 4$ eV,
and $\omega_C=7.2$ eV, in Fig.\ \ref{fig1}(b). Also shown is the measured
x-ray absorption spectrum (solid line). The final states of this latter
process are the intermediate states of the RIXS process. For example, the
broad peaks around 8991 eV and 8998 eV in the absorption spectrum have been
associated with the well-screened ($\underline{1s} 3d^{10} \underline{L}
4p$) and poorly-screened ($\underline{1s} 3d^9 4p$) core hole final states
\cite{Li91}, respectively, where $\underline{L}$ denotes the hole in oxygen
ligand. With this assignment, the intermediate states
responsible for the resonant enhancement of the low energy features (i.e. A
and B) are the well-screened states and those for the higher energy feature
(C), the poorly-screened states.

The low energy excitations A and B have relatively sharp resonance profiles.
As a result, one can see that the incident energies for which A and B
exhibit maxima differ slightly, by about 1 eV. It should be noted in Fig.\
\ref{fig1}(a) that due to matrix element effects, which depend strongly on
$E_i$, the spectral lineshape exhibits a strong $E_i$-dependence.  However,
the peak positions of A and B are largely independent of $E_i$. In the
rest of this paper, we focus on the {\bf Q}-dependence of these two
excitations, obtained with the incident photon energy fixed at $E_i=8991$
eV. A detailed analysis of the polarization and the incident energy
dependence will be presented elsewhere. We note, however, that when the
incident x-ray polarization is perpendicular to the {\bf c}-direction, the
behavior is qualitatively similar to that observed here, except for an
upward shift in the resonance energy by about 4 eV.

\begin{figure} \begin{center}
\epsfig{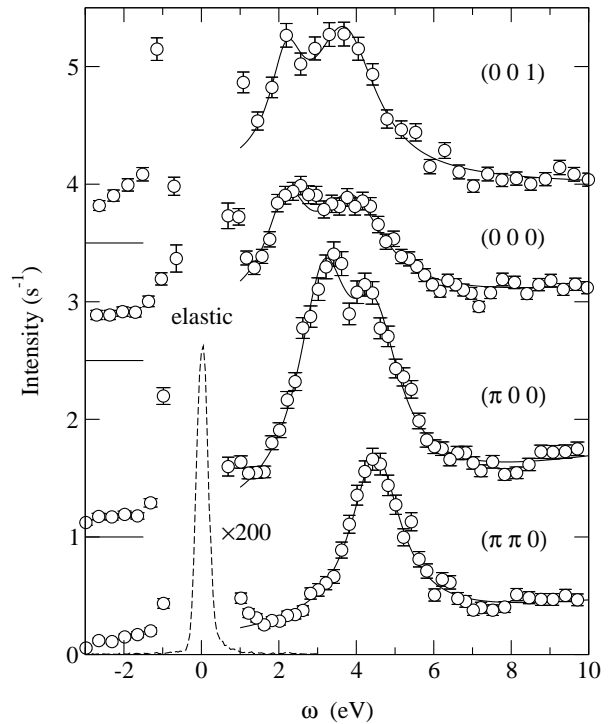}
\end{center} 
\caption{ 
RIXS spectra with $E_i=8991$ eV for a fixed reduced
wave vector ({\bf q}) as noted. Absolute momentum transfers ({\bf Q}),
respectively, 
correspond to (1 0 1), (1 0 0), (2.5 0 0), and (2.5 0.5 0), from top
to bottom. A representative scan through the elastic line is plotted to show
the instrumental energy resolution. Solid lines are fits to
a double Lorentzian lineshape as described in the text. The horizontal 
bars denote the baselines for each spectrum, which are offset
vertically for clarity. 
} 
\label{fig2}
\end{figure}

We have measured the energy spectra along the high-symmetry directions
{(0 0)-($\pi$ 0)} and {(0 0)-($\pi$ $\pi$)}, as well as the $L$-dependence
of the two-dimensional (2D) BZ center spectrum. Representative scans at the
high-symmetry positions, shown in Fig.\ \ref{fig2}, exhibit several
characteristics of the excitations: (i) The excitations A and B show clear
dispersion as {\bf q} is varied within the Cu-O plane, where {\bf q} is the 
reduced wavevector,
defined as ${\bf q} \equiv {\bf Q} - {\bf G}$, with {\bf G} denoting a
reciprocal lattice vector \cite{endnote}. (ii) They exhibit
dispersionless behavior along the $L$-direction, as shown in top two scans
in Fig.\ \ref{fig2}, and therefore can be regarded as effectively
two-dimensional. (iii) There is a dramatic change in the lineshape and/or
width at the ($\pi$ $\pi$) position. The double peak (A and B) feature is
still present at ($\pi$ 0), while only one peak is evident at ($\pi$ $\pi$).
One can fit the ($\pi$ $\pi$) data with a single Lorentzian with 0.8(1)
eV width.

\begin{figure}
\begin{center}
\epsfig{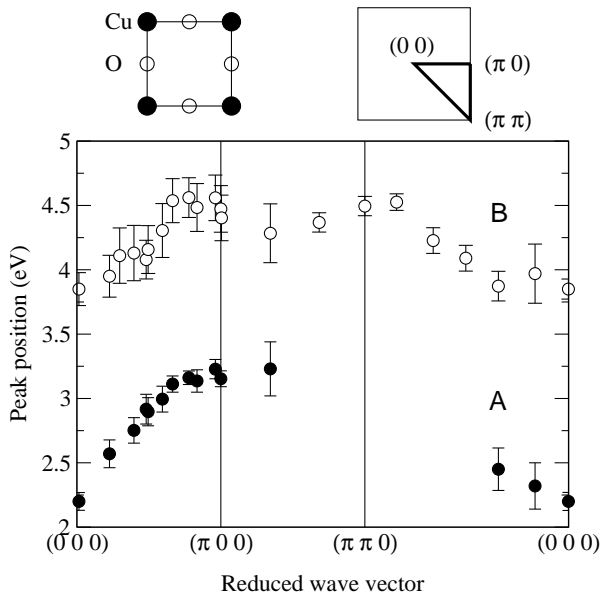}
\end{center}
\caption{
The peak positions of the double peak feature (A and B) obtained from the 
fitting are
plotted along the high symmetry directions. The
2D unit cell of the copper oxide plane and the 2D BZ are shown at the top.
}
\label{fig3}
\end{figure}

Away from the anomalous ($\pi$ $\pi$) position, we were able to analyze our data by
fitting to two Lorentzians. Our fitting results suggest that it is peak A that
becomes weaker and disappears at the ($\pi$ $\pi$) position. The results of these
fits are shown as solid lines in Fig.\ \ref{fig2}, and the fitted peak position is
plotted as a function of {\bf q} in Fig.\ \ref{fig3}. We note that it is very
difficult to extract reliable peak widths from fitting two overlapping peaks. In
addition, the peak intensity is strongly dependent on $E_i$. Thus, we focus on the
peak position as the most meaningful quantity to be extracted. Since both peaks are
much broader than our resolution, these results are insensitive to any resolution
effects, nor do they depend on the particular lineshape chosen. Furthermore, we
emphasize that the resolution is independent of the momentum transfer, so that the
unusual behavior around ($\pi$ $\pi$) cannot be a resolution effect.

As shown in Fig.\ \ref{fig3}, excitation A has a gap of 2.2 eV and a
dispersion of 1.0 eV, and it disappears near the ($\pi$ $\pi$) position. The
peak position of 2.2 eV at the zone center agrees well with that observed in
optical conductivity studies \cite{Uchida91,Falck92}. In contrast,
excitation B has a zone-center energy of 3.9 eV and a dispersion of 0.5 eV,
and does not exhibit strong momentum dependence to the intensity. Note that both 
excitations have a direct gap, that is, a minimum energy at the zone center.
This observation of {\bf q}-dependent excitations in the $2 \sim 4$ eV range is
broadly consistent with the recent RIXS studies of Ca$_2$CuO$_2$Cl$_2$
\cite{Hasan00} and of Sr$_2$CuO$_2$Cl$_2$ \cite{Abbamonte99}, although in
these experiments, the coarse energy resolution prevented the authors from
being able to resolve any double peak features. In addition, the amount of
dispersion observed here along the Cu-O bond direction ($\sim 1.0$ eV) is
almost twice that observed for Ca$_2$CuO$_2$Cl$_2$ and in fact is close to
that observed in the quasi-one-dimensional cuprate Sr$_2$CuO$_3$
\cite{Hasan02}. The data in Fig.\ \ref{fig3} also show some similarities to
the EELS results of Wang {\it et al.} on Sr$_2$CuO$_2$Cl$_2$ \cite{Wang96},
for which they observed two excitations with zone-center energies of 2.5 eV
and 4.1 eV, which disperse with large intensity changes as the momentum is
varied. Wang and coworkers identified the 2.5 eV mode as a CT exciton mode
with $p$-wave symmetry.

This assignment of the 2.5 eV feature as an exciton is consistent with a
number of optical conductivity studies of layered copper oxides, which have
associated a sharp peak near the optical gap ($\sim 2.2$ eV for
La$_2$CuO$_4$) with a bound exciton mode \cite{Uchida91,Lovenich01}.
However, this is not without controversy; Falck and coworkers have argued
that the sharp feature in the optical data arises from the short-range
interaction between an {\it unbound} electron and hole pair created by the
CT process \cite{Falck92,Thio90b}.  The present data, and in particular the
{\bf q}-dependence of Fig.\ \ref{fig3}, provide new insight into this
question, and, as discussed below, place stringent constraints on any
quantitative theories of the CT excitation spectrum.

First, the minimum of the excitation energy spectrum is located at the 2D BZ center.
This immediately highlights the importance of the strong electron correlations in
determining the excitation gap, since in the absence of such correlations simple
interband transitions would be expected to have a minimum excitation energy at
($\pi/2$ $\pi/2$). This is a result of the fact that the ZR band maximum is located
at ($\pi/2$ $\pi/2$) \cite{Wells95}, while the upper Hubbard band minimum is at
($\pi$ 0) \cite{Tsutsui99}. Calculations of two-particle excitations which include
the effect of correlations -- in the context of the so-called {\it t-t'-t"-U} model
\cite{Tsutsui99} -- predict that such excitations in fact have a minimum at ($\pi/2$
0), in disagreement with our observations. Second, as pointed out in Ref.\
\cite{Wang96}, the dispersion of excitation A ($\sim 1$ eV) is much larger than the
bandwidth of a single hole in the ZR band \cite{Wells95}, which suggests that
electrons and holes form a composite (i.e., bound) object that disperses more easily
\cite{Zhang98,Wang96}. Third, the strong {\bf q}-dependence of the intensity suggests
that the symmetry -- of a bound exciton -- may play an important role. Specifically,
for ${\bf q} \neq 0$, the symmetry of any exciton eigenstates do not remain pure, but
become mixed with states of other symmetries, for which different selection rules
apply \cite{Zhang98,Moskvin02}. Taken together, the above arguments suggest that the
behavior of the 2.2 eV feature is consistent with a bound exciton state.

Turning to the origin of excitation B, we first note that this excitation also shows
a strong resonance behavior and that therefore excitations involving the La bands can
be ruled out. One possibility then is that this excitation is in fact the broad
continuum arising from interband transitions creating electron-hole pairs. This
continuum is known to begin around 2 eV from photoconductivity data \cite{Thio90b}.
In this scenario, the observed {\bf q}-dependent lineshape changes would reflect
changes in the band structure.

However, it is more likely that excitation B is in fact an additional
exciton mode. First, in common with
the 2.2 eV feature, the excitation has a minimum at the
zone center, exhibits dispersion larger than that of a single hole and has a
Lorentzian lineshape -- the latter suggestive of a finite lifetime
excitation. In addition, the RIXS cross-section emphasizes local excitations over delocalized
ones, since the intermediate state of the resonance process is
itself spatially compact and therefore has a much larger overlap with a
localized final state.  Specifically, the
$\underline{1s}3d^{10}\underline{L}4p$ intermediate state involved here will
have a large overlap with a CT exciton \cite{Ide00} and therefore strongly enhance 
the
inelastic scattering from such objects, relative to that from the continuum
of delocalized electrons and holes. In optical measurements,
L{\"o}venich {\em et al.} estimated that the spectral weight of the exciton
was approximately twice that of the continuum \cite{Lovenich01}.  If this
ratio is indeed significantly enhanced for RIXS then -- based on the count
rates of the 2.2 eV feature, the continuum would be expected to be at the
limit of detection. Based on this and the above physical arguments, we
conclude that excitation B is also a CT exciton.

There are at least two possibilities for the nature of this higher-energy
exciton mode. One is that the hole again resides in the ZR band, 
but has
a different symmetry to that of the 2.2 eV feature. For example, Wang {\em
et al.} have argued that the 4.1 eV feature observed in EELS work on
Sr$_2$CuO$_2$Cl$_2$ is the same CT exciton but with $s$-wave symmetry
\cite{Wang96}. Another possibility is that the hole is in fact in the
non-bonding oxygen $p_{\pi}$ orbitals on the same CuO$_4$ plaquette as the
electron \cite{Pothuizen97,Choi99}. Unfortunately, our data do not
distinguish between these two possibilities and detailed theoretical
calculations will be required to resolve this question.

The picture our data present then is of two distinct, highly dispersive
exciton modes in La$_2$CuO$_4$, both residing {\it in} the electron-hole
continuum.  The presence of the continuum provides numerous decay channels
and as a result the excitons are highly damped and the observed peaks are
broader than the resolution. These excitons are quite different from
conventional excitons, observed in semiconductors for example, and our data
point to the need for detailed theoretical calculations to explain their
properties, in particular the disperson relations and momentum-dependent
intensities. In turn such calculations will shed light on the charge
dynamics of the copper oxide layers.

In summary, we have carried out a RIXS study of the CT excitation spectra as a
function of {\bf q} along the high-symmetry directions in insulating La$_2$CuO$_4$.
We have observed two highly-dispersive and highly-damped exciton-like CT excitations,
both of which show two-dimensional characteristics.  The low energy mode has a gap of
2.2 eV and bandwidth of 1.0 eV, and shows a strong {\bf q}-dependent intensity
variation. The second peak shows a smaller dispersion ($\sim 0.5$ eV) with a
zone-center energy of $\sim 3.9$ eV.


We would like to thank F. Essler, P. D. Johnson, C.-C. Kao, M. A. Kastner, G. A. Sawatzky,
B. O. Wells, and F. C. Zhang for invaluable discussions. The work at Brookhaven was
supported by the U. S. Department of Energy, Division of Materials Science, under contract
No. DE-AC02-98CH10886. Use of the Advanced Photon Source was supported by the U. S.
Department of Energy, Basic Energy Sciences, Office of Science, under contract No.  
W-31-109-Eng-38. CAB was supported by the U. S. Department of Energy, Division of Materials
Science, under contract No. DE-FG02-99ER45772.


\begin{thebibliography}{10}
\bibitem{endnote}
We have verified that the dispersion relation
measured by RIXS depends only on {\protect \bf q}, not {\protect \bf Q}, by
measuring spectra at several {\protect \bf Q} positions with the same
{\protect \bf q}.
\end{thebibliography}
\end{document}